\documentclass{hep99}
\def \gam {\frac{ N_f N_cg^2_{\pi q\bar q}}{8\pi} }
\begin{document}

\title{The light scalars and the broad $\sigma(500)$
in the U3$\times$U3 linear sigma model\footnote{Talk given at the
International Conference on High Energy Physics (Hep99),  July 15-21,
1999 Tampere Finland}}

\author{N. A. T\"ornqvist}
%
% Use of footnote symbols, footnoted material after \maketitle 
%\author{A J Cox$^1$\dag\ and Jim Revill$^2$\ddag}

\address{
Physics Department, University of Helsinki, Helsinki, Finland\\
E-mail: {\tt Nils.Tornqvist@Helsinki.fi}}
\abstract{The lightest
scalar and pseudoscalar nonets are discussed within the framework
of the broken old 
U3$\times$U3 linear sigma model, and it is shown that already 
at the tree level this model
 works remarkably well predicting scalar masses and 
couplings not far  from present experimental values, 
when all parameters are fixed from the 
pseudoscalar masses and decay constants. The linear $\sigma$ model
is the simplest way to
implement chiral symmetry together with the broken 
SU3 of the quark model, and this, not well known, success in understanding 
experiment is comparable to that of
the naive quark model for the heavier multiplets. It is argued that this
strongly suggest that the light and very broad $\sigma$
resonance exists near 500 MeV.
} 

\maketitle

% Text of footnotes comes after \maketitle
%\fntext{1}{E-mail: tony.cox@ioppublishing.co.uk}
%\fntext{2}{E-mail: jim.revill@ioppublishing.co.uk}
%\fntext{\dag}{Here is a footnote.}

\def \gam {\frac{ N_f N_cg^2_{\pi q\bar q}}{8\pi} }
\def \gamm {N_f N_cg^2_{\pi q\bar q}/(8\pi) }
\def \be {\begin{equation}}
\def \ba {\begin{eqnarray}}
\def \ee {\end{equation}}
\def \ea {\end{eqnarray}}
\def \gap {{\rm gap}}
\def \gaps {{\rm {gaps}}}
\def \gappp {{\rm \overline{\overline{gap}}}}
\def \im {{\rm Im}}
\def \re {{\rm Re}}
\def \Tr {{\rm Tr}}
\def \P {$0^{-+}$}
\def \S {$0^{++}$}
\def \uu {$u\bar u+d\bar d$}
\def \ss {$s\bar s$}

%\se

 The lightest  scalars, the 
$a_0(980),$ $f_0(980),$ $ K^*_0(1430)$ and the $\sigma(400-1200)$,
which we shall here call $\sigma(500)$, have remained controversial for long, 
since the naive quark model, without chiral symmetry and finite widths from
unitarity, fails badly in trying to accomodate them. 
Today many authors want to give the 
$a_0(980),\ f_0(980)$ and the $\sigma(500) $ other interpretations 
than being $q\bar q$ states. Popular alternative interpretations are 
$K\bar K$ bound states, 4 quark states,
or for the $\sigma$, a glueball.

But, in fact, there is an old chiral quark model, the   
linear U3$\times$U3 sigma model in which one can 
treat both the scalar and pseudoscalar nonets simultaneously
with chiral symmetry. 
Unfortunately this over 30 years old 
model\cite{sigma} has had very few recent phenomenological 
applications, and therefore its success to qualitatively describe data 
has been forgotten.  

The  Lagrangian  is (for more details see \cite{NATEPJ}):  
\ba 
 {\cal L}&=&
\frac 1 2 \Tr [\partial_\mu\Sigma \partial_\mu\Sigma^\dagger]
-\frac 1 2 \mu^2\Tr [\Sigma \Sigma^\dagger]\nonumber \\
 &-&\lambda \Tr[\Sigma\Sigma^\dagger \label{lag} 
\Sigma\Sigma^\dagger]\ -\lambda' 
(\Tr[\Sigma\Sigma^\dagger])^2 \\
&+&\epsilon_\sigma \sigma_{u\bar u+d\bar d} + 
\epsilon_{s\bar s} \sigma_{s\bar s} +
\beta [\det \Sigma +\det \Sigma^\dagger] \ .\nonumber
\ea
Here the $\epsilon$ and $\beta$ terms 
give the pseudoscalars mass and  break the 
flavour and $U_A(1)$ symmetries. 
The stability condition, that the linear terms
in the fields must vanish after the shift of the scalar fields 
($\Sigma\to \Sigma +V$) determines
the small parameters $\epsilon_i$ in terms of 
the pion and kaon masses and decay constants. One finds 
 $\epsilon_\sigma = m_\pi^2f_\pi$, $\epsilon_{s\bar s}=
(2m_K^2f_K-m_\pi^2f_\pi)/\sqrt 2$, 
while $\beta$ in the $U_A(1)$ breaking term is 
determined by $m_{\eta '}$, or by 
$m^2_\eta+m^2_{\eta '}$.

My previous work on the scalars with the unitarized 
quark model (UQM)\cite{NAT} 
is essentially a  unitarization of 
eq.(\ref{lag}) with $\lambda\approx 16$ and $\lambda '= 0$, 
and with the main symmetry breaking generated by 
putting the pseudoscalar masses  at their physical values.

It is an ideal problem for a symbolic program like Maple V to calculate the
predicted masses, and couplings from the Lagrangian, which has 
6 parameters, $\mu,\lambda,\lambda',\beta$, $u=d$ and $s$, of which the
last two define the diagonal 
matrix $V$ with the flavourless meson VEV's: $V=diag[u,d,s]$. These are at the 
tree level related to the pion and kaon decay constants through
 $u=d=<\sigma_{u\bar u,d\bar d}>/\sqrt 2=f_\pi/\sqrt 2$ (assuming 
isospin exact) and $s=<\sigma_{s\bar s}>=(2f_K-f_\pi)/\sqrt 2$.
One finds  denoting the often occurring combination 
$\mu^2+4\lambda'(u^2+d^2+s^2)$ by $\bar \mu^2$:
\ba
m^2_{\pi^+}\ &=&\bar \mu^2 + 4\lambda(u^2+d^2-ud)+2\beta s\ , \\
m^2_{K^+}  \ &=&\bar \mu^2 + 4\lambda(u^2+s^2-su)+2\beta d\ , \\
m^2_{a_0^+}\   &=&\bar \mu^2 + 4\lambda(u^2+d^2+ud)-2\beta s \ ,\\
m^2_{\kappa^+} \  &=&\bar \mu^2 + 4\lambda(u^2+s^2+su)-2\beta d \ .
\ea
For the masses and mixings of isoscalar states see Ref\cite{NATEPJ}.

 We can fix 5 of the 6 parameters, leaving  $\lambda'$ free,
 by the 5 experimental 
quantities from the pseudoscalar sector alone:
$m_\pi$, $m_K$, $m^2_\eta+m^2_{\eta'}$,
$f_\pi=92.42$ MeV and $f_K=113$ MeV, which all are accurately known from experiment. One finds that at the tree level 
$£ \lambda=11.57,\  
\bar \mu^2=0.1424\ {\rm GeV}^2,$ $ \beta =-1701\ {\rm MeV, }$$
\ u=d=65.35\ {\rm MeV},\ $$ s=94.45\ {\rm MeV}.$
The remaining $\lambda'$ parameter changes only the $\sigma$ and $f_0$ masses 
and their trilinear couplings, 
not those of the pseudoscalars. 
It turns out that $\lambda '$ must be small,
compared to $\lambda$, in order to fit the tri-linear couplings. 
By putting $\lambda'=1$ one gets a reasonable compromise for most of 
these couplings. With $\lambda'\approx 3.75$ one almost 
cancels the OZI rule breaking 
coming from the determinant  term, and the scalar mixing becomes near ideal
(for $\lambda '=-\beta /(4s)=4.5$ the cancellation is  exact).
   
\begin{table}[t]
\centering
\caption{ \it Predicted masses in MeV and mixing angles for two values of the 
 $\lambda'$ parameter. The 
 asterix means that $m_\pi,m_K$
and $m_\eta^2+m^2_{\eta'}$ are fixed by experiment together with $f_\pi=$92.42 MeV and 
$f_K=$113 MeV.  }
\vskip 0.1 in
\begin{tabular}{|l|c|c|c|} \hline
Quantity       &  Model $\lambda '=1$ & Experiment \\
\hline
\hline
$m_\pi$   &  137$^{*)}   $ &137  \\
$m_K$     &  495$^{*)}   $ &495  \\
$m_\eta  $&  538$^{*)}   $ &547.3  \\
$m_{\eta'}$ &963$^{*)}   $ &957.8  \\ 
$\Theta^{\eta'-singlet}$ &-5.0$^\circ$ &(-16.0$\pm6.5)^\circ$ \\ 
$m_{a_0}$ &1028 &  983  \\ 
$m_{\kappa}$ &1123 &   1430  \\ 
$m_{\sigma}$ &651 & 400-1200  \\ 
$m_{f_0}$ &1229    &980  \\ 
$\Theta^{\sigma-singlet}$ 
& 21.9$^\circ$  &(28-i8.5)$^\circ$\cite{NAT}  \\
\hline
\end{tabular}
\label{tab1}
\end{table}

As can be seen from Table 1 the predictions 
 are not far from the experimental masses taken as $a_0(980),\ f_0(980),\
K^*_0(1430)$, and $\sigma(500)$. In particular note that one predicts
a low mass for the controversial  \uu\ scalar meson of 650 MeV, and which
as we shall see should have a very large width (Tables 2-3 below). This is
essentially a zero parameter prediction once the main parameters are fixed 
from the data on pseudoscalars. 
Considering that one expects  that unitarity 
corrections can  be up to 30\%, and should 
go in the right direction compared to experiment, one must
conclude that these results for the other scalar masses ($a_0,f_0(980)$ 
and $K^*_0$) are good enough to take the model seriously. 

The trilinear coupling constants follow from the Lagrangian after 
one has made the shift $\Sigma\to\Sigma+V$. The predicted
spp couplings at the tree-level can be  expressed
in terms of the predicted physical masses and mixing
angles and decay constants. E.g.:
\ba
g_{\kappa^+K^0\pi^+}&=&(m^2_\kappa-m^2_\pi)/(\sqrt 2 f_K)\ , \\  
g_{\sigma\pi^+\pi^-}&=&\cos \phi^{s\bar s-f_0}(m^2_\sigma-m^2_\pi)/f_\pi\ , \\
g_{f_0\pi^+\pi^-}&=&\sin \phi^{s\bar s-f_0}(m^2_{f_0}-m^2_\pi)/f_\pi \ ,
 \label{f0pipi}\\
g_{a_0\pi\eta}&=&\cos \phi^{s\bar s-\eta '}(m^2_{a_0}-m^2_\eta)/f_\pi \ ,\\
g_{a_0\pi\eta'}&=&\sin \phi^{s\bar s-\eta '}(m^2_{a_0}-m^2_{\eta'})/f_\pi
 \ ,\\  
g_{a_0 K^+K^-}&=&(m^2_{a_0}-m^2_K)/(2f_K) \ .
\ea

For more predictions see Ref.\cite{NATEPJ}. 
In Table 2 several different spp couplings are compared with quoted 
experimental numbers. 

\begin{table}[t]
\centering
\caption{ \it Predicted couplings $\sum_i\frac{g_i^2}{4\pi}$ (in GeV$^2$)
, when $\lambda'=1$, 
compared with experiment\cite{pdg98,aston,ach,nov}. (We 
have used isospin invariance to get the sum over charge channels, when there
is data for one channel only.)    The  
$a_0\pi\eta$ coupling 
is very sensitive to loop corrections due to the $K\bar K$ threshold.}
\vskip 0.1 in
\begin{tabular}{|l|c|c|c|c|} \hline
Process       &  $ \sum_i\frac{g_i^2}{4\pi}$ &  $ \sum_i\frac{g_i^2}{4\pi}$   \\
      & in model &in experiment \\ 
\hline
\hline
$\kappa^+\to K\pi$   &  7.22 & -    \\
$\kappa^+\to K^+\eta $     &  0.28 & $\approx 0$ \\
$\sigma\to \pi\pi$   &  2.17 & 1.95  \\
$\sigma\to K\bar K$   &  0.16 & 0.004  \\
$f_0\to    \pi\pi$   
&  1.67 & 0.765$^{+0.20}_{-0.14}$ \\
$f_0\to    K\bar K$   &  6.54 & 4.26$^{1.78}_{-1.12}$ \\
$a_0^+\to \pi^+\eta$   &  2.29 & 0.57  \\
$a_0^+\to K\bar K$ &  2.05 &  1.34$^{+0.36}_{-0.28}$  \\
\hline
\end{tabular}
\label{tab2}
\end{table}
\begin{table}[t]
\centering
\caption{ \it Predicted widths compared with experiment\cite{pdg98,
aston,ach,nov} 
(in MeV).   The 
predicted $f_0\to \pi\pi$ width is extremely
 sensitive to the value of $\lambda^\prime$ (for $\lambda '= 
3.75 $  it nearly vanishes)
and unitarity effects. Also the  
$a_0\pi\eta$ coupling 
is very sensitive to loop corrections due to the $K\bar K$ threshold.}
\vskip 0.1 in
\begin{tabular}{|l|c|c|c|c|} \hline
Process         &
$\sum_i\Gamma_i$ & $\sum_i\Gamma_i$ \\
       & model&experiment\\ 
\hline
\hline
$\kappa^+\to K\pi$  & 678 &  $278\pm23$   \\
$\kappa^+\to K^+\eta $     & 13 & $<26$ \\
$\sigma\to \pi\pi$    & 574  &300-1000 \\
$f_0\to    \pi\pi$    & see text & 40 - 100\\
$a_0^+\to \pi^+\eta $& 273 see text & 50 - 100\\
\hline
\end{tabular}
\label{tab3}
\end{table}

As can be seen from Tables 2-3 most of the  couplings 
are not far from experiment.
Only the $f_0\to\pi\pi$ and $a_0\to\pi\eta$ couplings and widths
come out a bit large, but these are very sensitive to higher order loop  
corrections due to the $K\bar K$ threshold, and $f_0\to\pi\pi$ is 
extremely sensitive to the scalar near-ideal mixing angle and $\lambda '$.
 If one choses $\lambda'= 3.75$  this mixing angle nearly vanishes 
($\phi^{s\bar s-f_0}=-3.0^\circ$) together with the $f_0\to\pi\pi$ coupling.
From our experience with the UQM\cite{NAT} the $a_0\to K\bar K$ peak width, 
when unitarized, is reduced, because of
 the $K\bar K$ theshold, by up to a factor 5. Therefore one cannot 
expect that the tree level couplings should agree better with data 
than what those of Tables 2-3 do.  After all, this is a very strong
coupling model ($\lambda=11.57$, leading to large $g_i^2/4\pi$)
and higher order effects should be important.

In summary, I find that the linear sigma model 
with three flavours, at the tree level, works much  
better than what is generally believed.
 When the 6 model parameters are fixed mainly by the
pseudoscalar masses and decay constants,
 one predicts the 4 scalar masses
and mixing angle to be reasonably near those of the experimentally observed nonet $
a_0(980),$ $ f_0(980),$ $\sigma(500),$ $ K^*_0(1430)$.
Also 8 couplings/widths of the scalars to two pseudoscalars are predicted
reasonably close to  their  presently known, 
rather uncertain experimental values. The agreement is good enough considering
that 
some of these are expected to have large higher order corrections. The model
 works, in my opinion, just as well as the 
naive quark model works for the heavier nonets.
A more detailed data comparison would become meaningful,
after one  has included  higher order effects, i.e. after one 
has  unitarized the model, 
e.g., along the lines of the UQM\cite{NAT}. 
Of course, we believe by no means that the sigma model is a fundamental theory,
only that it is 
a reasonable  effective theory at low energy, which in a compact way can
incorporate constraints from symmetry and symmetry breaking. 
With only a few low-dimensional invariants, like in Eq.(1), 
the model is renormalizable,
i.e. it can provide a good starting point for the inclusion of
 unitarity and analyticity effects.  

Those working on chiral perturbation theory and nonlinear sigma models
 usually point out that the linear model does not predict all low energy 
constants correctly. However, one should remember that the energy regions 
of validity are different for the two approaches.
 Chiral perturbation theory usually breaks down when one approaches
the first scalar resonance. 
The  linear sigma model, on the other hand,
includes the scalars from the start 
and can be a reasonable interpolating model in the intermediate 
energy region near 1 GeV, where QCD is too difficult to solve.

These results strongly favour the interpretation that the 
$a_0(980)$, $f_0(980)$, $\sigma(500)$,  $ K^*_0(1430)$  
belong to the same nonet, 
and that they are the chiral partners of the $\pi$, $\eta$,  $\eta '$, $ K$. 
If the latter are believed to be unitarized $q\bar q$ states, 
so are the light scalars  $
a_0(980),$ $ f_0(980),$ $\sigma(500),$ $ K^*_0(1430)$, and the broad 
$\sigma(500)$ should be interpreted as an existing resonance.
The $\sigma$ is a very important hadron indeed, 
as is evident in the sigma model, because this is the boson
which gives the constituent quarks most of their mass
and thereby it gives also 
the light hadrons most of  their mass. Therefore it is natural to consider
the $\sigma (500)$ as the Higgs boson of strong interactions. 

{\it Acknowledgement.} This work is partially supported by
 the EEC-TMR program Contract N.CT98-0169.

\end{document}